\documentclass[preprint,aps,nofootinbib]{revtex4}
\usepackage{graphicx}
\input{epsf.sty}
\usepackage{epsfig}

\begin{document}

\def\err#1#2{\stackrel{\scriptstyle +#1}{\scriptstyle -#2}}
\def\gev{\rm GeV}
\def\tev{\rm TeV}
\def\fbi{\rm fb^{-1}}
\newcommand{\ms}{{M_S^{}}}
\newcommand{\mssq}{{M_S^2}}
\newcommand{\str}{string}
\def\lsim{\mathrel{\raise.3ex\hbox{$<$\kern-.75em\lower1ex\hbox{$\sim$}}}}
\def\gsim{\mathrel{\raise.3ex\hbox{$>$\kern-.75em\lower1ex\hbox{$\sim$}}}}

\hfill$\vcenter{\hbox{\bf MADPH-03-1343}
                \hbox{\bf IC/2003/64}
                 \hbox{\bf hep-ph/0309132}}$
\vskip 0.4cm

\title{Bounds on Four-Fermion Contact Interactions\\  Induced by 
String Resonances}
\author{P. Burikham$^1$\footnote{piyabut@pheno.physics.wisc.edu}, 
T. Han$^1$\footnote{than@pheno.physics.wisc.edu}, 
F. Hussain$^2$\footnote{hussainf@ictp.trieste.it}, 
and D. W. McKay$^3$\footnote{mckay@kuark.phsx.ukans.edu} }

\affiliation{$^1$Department of Physics, 1150 University Ave., 
University of Wisconsin, Madison, WI 53706, USA}
\affiliation{$^2$The Abdus Salam International Center for Theoretical, 
Physics, Trieste, Italy}
\affiliation{$^3$
Department of Physics \& Astronomy, University of Kansas, 
Lawrence, KS 66045, USA}

\date{\today}

\vskip 1cm

\begin{abstract}
Based on tree-level open-string scattering amplitudes in the low
string-scale scenario, we derive the massless fermion
scattering amplitudes.  The amplitudes are required to reproduce
those of the Standard Model at tree level in the low energy limit.  
We then obtain four-fermion contact interactions by expanding in inverse
powers of the string scale and explore the constraints on the string
scale from low energy data.
The Chan-Paton factors and the string scale are treated as free parameters.
We find that data from the neutral and charged current processes at HERA, 
Drell-Yan process at the Tevatron, and from LEP-II put lower bounds 
on the string scale $M_S$, for typical values of the 
Chan-Paton factors, in the range $M_S \geq 0.9-1.3$ TeV, comparable to Tevatron bounds on $Z^\prime$ and $W^\prime$ masses.
 
\end{abstract}


\maketitle

\section{introduction}

String theory, spoken or unspoken, is generally assumed to be the
underpinning of the low scale gravity ideas \cite{add,RS1} explored
theoretically and experimentally in recent years. A number of examples
of ambitious ``top-down'' models of string realizations of low scale
gravity ideas have been advanced, aiming at consistently achieving the
connection to Standard Model (SM) physics from higher mass scales in
certain D-brane scenarios \cite{stringy}.  As yet a fully realistic
model like the SM has not been constructed.  On the other hand, one
could take a more phenomenological approach, from the ``bottom
up''. One of the recent endeavors is to obtain the SM tree-level
amplitudes at low energies \cite{cpp,cim,fhh} based on open-string
amplitudes \cite{aki,pol}. This approach assures the correct low
energy phenomenology as given by the SM, yet captures one of the essential features
of string theory, namely the string resonances, in a relatively
model-independent way.  The basic assumptions in this approach are
that the fundamental string scale $\ms$ is at the order of one $\tev$,
and that the dominant contributions to the low energy processes are
due to the exchange of string resonances.  Earlier work on
phenomenological studies dealt with QED from the $M_Z$ scale to the
first few string resonances \cite{cpp}, or neutrino inclusive
processes far above the string scale to explore the effects from
cosmic neutrinos \cite{cim,fhh}. Phenomenologically, this
string-amplitude approach complements the low-scale gravity 
calculations based on expansions in Kaluza-Klein modes \cite{extrad},
which are argued to be higher orders in string-coupling 
expansion \cite{stringy,cpp}.

The purpose of this note is to expand this effort to model both {\it
neutral and charged current} interactions at energies below the string
scale.
 
Data from HERA experiments at DESY, with lepton-parton center of mass
(CM) energies receiving a good fraction of the full 320 GeV, the
highest energy available in laboratory experiments for deep inelastic
scattering, provides one interesting testing ground for low-scale
string model ideas. Similarly LEP-II, with CM energies up to 200 GeV
provide another reasonably sensitive probe of the low energy
limit of our string-resonance amplitudes.  The full CM energy is
available to excite string effects in this case.  At the Tevatron,
though the parton-parton collisions get typically only a modest
fraction of the $1.8-2$ TeV available in the $p\overline p$ CM energy, 
there is still sensitivity to $0.5 - 1$  TeV scale physics.

The good agreement between all of the data from the facilities just
mentioned and the SM allows bounds to be set on the mass scale of all
kinds of new physics effects.  For example, leptoquark states are one
such effect, and perturbative string resonances can carry the same
quantum numbers as the lepto-quarks in some channels \cite{cim,fhh}.
At the parton level, much of the kinematical range is low enough to
justify keeping the lowest order terms in an expansion in inverse
string scale. This allows a direct comparison of amplitudes with the
existing limits on new physics contact interactions
\cite{bchz,kingman,PDG,peskin}.  These observations that a
comprehensive bound can be applied to a wide class of string-resonance
models motivate the work we present here. We hope that exploration of
the constraints imposed on the model parameters 
by the agreement between data and the SM will ultimately shed light on
the way string theory signals could emerge as laboratory energies rise
above the currently available regime.

This paper is organized as follows. In Section \ref{Amplitudes}, we
summarize the construction of neutral and charged current interactions
for SM light fermions based on open-string scattering amplitudes. We
then in Sec.~\ref{contact} take the low-energy expansion 
by expanding the string amplitudes in powers of the inverse string scale
evaluated at typical kinematic points to obtain the effective
four-fermion contact-like interactions. We check that the approximation is
good in the kinematic ranges we use. Comparing with the current limits on
these interactions, we derive bounds on the string scale $\ms$. We
conclude in Sec.~\ref{conclude}. 

\section{Open String Tree Graph Amplitudes} 
\label{Amplitudes}

In weakly-coupled string theory with a low string-scale, one generically 
expects the string amplitude corrections to the standard model processes 
to dominate over the graviton corrections, which enter at one loop and 
are parameterically suppressed by an
extra factor of $g^2$, a gauge coupling squared \cite{stringy,cpp,pol}.
At energies well below $\ms$, the stringy corrections can
be systematically taken into account by the low-energy expansion of
the string amplitudes in terms of $s/\mssq$. 

We assume that the tree-level string amplitudes represent the
scattering of massless SM particles, as the zero string modes. The
first attempt at exploring the low-scale string amplitudes was made
to construct a string toy model of QED of electrons and
photons \cite{cpp}. The SM is embedded in a type IIB string theory
whose 10-dimensional space has six dimensions compactified on a torus
with common period $2\pi R$.  There are N coincident D3-branes, on
which open strings may end, that lie in the 4 extended dimensions.
The extra symmetry of the massless string modes are eliminated by
(unspecified) orbifold projection.  The paper applies the results to
Bhaba scattering and then adds several prescriptions to include some
simple processes $e^{+} e^{-} \rightarrow Z^0 \rightarrow e^{+} e^{-}$
and $q \bar{q} \rightarrow g^* \rightarrow$ 2 jets, where $g^*$
represents a string resonance excitation of the gluon.  However, this
toy model does not attempt to be fully realistic in terms of the SM
particle spectrum and their interactions.

Our construction of the tree graph amplitude follows the same pattern
as that outlined in \cite{cim} and in \cite{fhh}.  The result is a
model containing the SM on the 3-branes and no unacceptable ({\it
i.e.,} unobserved) low energy degrees of freedom. This is accomplished
by allowing the group theoretical Chan-Paton factors as free
parameters.  The masses of gauge bosons $W$ and $Z$ must be introduced
by hand, since the string amplitude describes massless particle
scattering and we are not consistently modeling the breaking of gauge
invariance.  Though all the standard model gauge couplings are assumed to
unify to a single value at the string scale in this simple construction, we
use the physical values of the SM electroweak couplings since we
restrict ourselves here to energies below the string scale.

We begin with the general form for a four-fermion amplitude for open
strings in such a braneworld framework.  The parton level Mandelstam
variables are denoted by $s,t,$ and $ u$.  The physical scattering
process will be identified as $f_{1}+f_{2} \rightarrow f_{3}+f_{4}$.
The $s, t$ and $u$-channels are labeled (1,2), (1,4) and (1,3),
respectively.  The ordered amplitude with the convention that all
momenta are directed inward reads \cite{pol,fs,js}:
\begin{eqnarray*}
A_{\str}(s,t,u)&= & ig^2  \left[t\ \overline{\psi}_{1}\gamma^{\mu}\psi_{2}\overline{\psi}_{3}\gamma_{\mu}\psi_{4} - s\ \overline{\psi}_{1}\gamma^{\mu}\psi_{4}\overline{\psi_{3}}\gamma_{\mu}
\psi_{2}\right] \\ & \times & \left[\frac{S(s,t)}{st} [T(1234)+T(4321) ] +(1\leftrightarrow 4,
s\leftrightarrow u) +(1\leftrightarrow 2, t\leftrightarrow u)\right].
\end{eqnarray*}
where the function $S(x,y)$ is similar to a Veneziano amplitude \cite{gv}, and is defined by
\begin{equation}
S(x,y) = \frac{\Gamma(1-\alpha ' x) \Gamma(1-\alpha ' y)}{\Gamma(1-\alpha 'x
-\alpha ' y)},
\end{equation}
where the Regge slope parameter $\alpha ' = M_S^{-2}$.  In the limit
$\ms \gg \sqrt s$, $S \rightarrow 1$ and the low energy gauge theory
expression for the amplitude is regained, as we show below.  The
factors $T(1234)+T(4321)$ and their $1\leftrightarrow 4$ and $
1\leftrightarrow 2$ counterparts are proportional to the Chan-Paton
factors \cite{cp} and involve traces over the group representation
matrices, $\lambda$, of the fermions at the four vertices.  For
example, $T(1234) \propto Tr(\lambda^1\lambda^2\lambda^3\lambda^4)$
with normalization $Tr(\lambda^a\lambda^b)=\delta^{ab}$ in the adjoint representation of $U(n)$.  Typically, with our normalization, the Chan-Paton factors are in the range of
$-4$ to $4$ for a general $U(n)$ group.  The above general expression
serves as the basis for calculating all of the specific helicity and
internal quantum number possibilities in the case that the states 3
and 4 have outgoing momenta.

\subsection{Charged Current Processes}

The charged current (CC) string model amplitude in the weak coupling
regime receives no contribution from the graviton at one loop.  In
this sense it is perhaps conceptually cleaner than the neutral current
(NC) case \cite{cim,fhh}, where the graviton exchange is contained in
the one loop amplitude \cite{cpp}.  At energies above the string
scale, the extra power of $s/\mssq$ in the graviton contribution
compensates for the Yang-Mills gauge coupling suppression of the loop
amplitude compared to the tree graphs; there the strong gravity dynamics
and the string resonance dynamics become comparable.  Though we are
focusing on the low energy region, where graviton exchange is
suppressed, the CC amplitude construction is simpler than that of the
NC because there are fewer processes and only one gauge coupling to
consider.  For this reason we discuss the CC case first in some
detail, and then turn to the NC case.

For definitiveness, taking all helicities for the in and out states
left-handed (denoted by $L$), we find the string tree amplitude:
\begin{eqnarray*}
A_{\str}^{CC}(LL) =  ig^2\left[ S(s,t) \frac{s}{t} T_{1234}
+ S(u,t)(- \frac{s}{t} - \frac{s}{u})T_{1324}
+ S(s,u) \frac{s}{u} T_{1243}\right],
\end{eqnarray*}
where we have further simplified notation by introducing $T_{1234} =
T(1234) + T(4321)$ and so forth.  The corresponding standard model
electroweak (EW) tree amplitude is
\begin{equation}
A_{EW}^{CC} = ig^2\frac{s}{t-M_{W}^2}.
\label{W}
\end{equation}
here and henceforth, $g$ is identified with the $SU(2)_{L}$ gauge coupling.
We require that the charged-current in $t$-channel contain the $W$ boson as
its zero mode and that there is no exotic (leptoquark) zero mode in
the $u$-channel. In order to remove the unwanted zero-mode pole, we must require
\begin{equation}
 T_{1243} = T_{1324} \equiv T.
\label{eq:2}
\end{equation}
The low energy gauge theory limit should reproduce the $W$-pole in the
$t$-channel in  tree approximation to the string amplitude.   
Using Eq.~(\ref{eq:2}) and matching the coefficient of the $1/t$ pole to 
the SM result of Eq.~(\ref{W}),  we identify
\begin{equation}
                   T_{1234} = 1 + T
\end{equation}
The tree-level result for the amplitude for $LL\rightarrow LL$ after 
removing the exotic zero-mode pole in the $u$-channel and identifying the
zero-mode pole in the $t$-channel as the $W$-boson, is
\begin{equation}
   A_{\str}^{CC}(LL)=ig^2 T\frac{s}{ut} f(s,t,u) + ig^2\frac{s}{t-M_{W}^2}S(s,t),
\label{eq:4}
\end{equation}
where
\begin{equation}
 f(s,t,u) \equiv uS(s,t)+sS(u,t)+tS(s,u).
\label{eq:60}
\end{equation}
In the limit $\ms \gg \sqrt{s}$, we have 
\begin{equation}
S(s,t)\approx 1- \frac{\pi^2}{6}\ \frac{st}{M^4_{S}} \quad {\rm and}\quad 
 f(s,t,u) \approx -\frac{\pi^2}{2}\frac{stu}{M_{S}^4}.
\label{eq:6}
\end{equation}
The SM tree amplitude is reproduced in the limit $s/M_S^2 \rightarrow 0$.
For later convenience we define $V(s,t,u)$ via Eq.~(\ref{eq:4}) by
\begin{equation}
A^{CC}_{\str}(LL) = ig^2\frac{s}{t-M_{W}^2}V(s,t,u).
\label{eqV}
\end{equation}

The above results are also applicable to right-handed anti-fermion
scattering $\overline R\ \overline R\to \overline R\ \overline R$.
The other helicity combinations including anti-leptons and anti-quarks
can be worked out by appropriate crossing.  For instance, for the
scattering of a left-handed lepton and a right-handed anti-quark
$L\overline{R}\rightarrow L\overline{R}$ , or right-handed anti-lepton
on left-handed quark $\overline{R} {L}\rightarrow \overline{R} L$, the
$s \leftrightarrow u$ and $2 \leftrightarrow 3$ crossed amplitude
applies.  The amplitudes for this process read
\begin{eqnarray}
\nonumber
A_{\str}^{CC}(L\overline{R}) &=& A_{\str}^{CC}(\overline{R} L)
  =  ig^2\left[ S(s,t) (- \frac{u}{t} - \frac{u}{s}) T_{1234}
   \mbox{}+ S(t,u)\frac{u}{t}T_{1324}
   \mbox{}+ S(s,u) \frac{u}{s} T_{1243} \right]\\
 &=&  A_{\str}^{CC}(LL)(s\leftrightarrow u) =
 ig^2\frac{u}{t-M_{W}^2}\overline{V}(s,t,u),
\label{eqVb}
\end{eqnarray}
where the generic label $T$ is not distinguished from that in the $LL$
case above, to avoid clutter in the notation.  These expressions are
the analogs of those written down for the NC neutrino case in
\cite{cim} and \cite{fhh}, which we expand for the full range of NC
cases in the next subsection.

\subsection{Neutral Current Processes}

The open string perturbative amplitude construction for $2 \rightarrow
2$ NC scattering follows exactly the same pattern as described above
in the charged current case.  The neutral current case involves
4-fermion amplitudes as well as 2-lepton plus 2-gluon external line
amplitudes \cite{fhh}.  We find that in the low energy realm the gluon
amplitudes contribute negligibly to the constraints. Therefore, we
confine ourselves to the 4-fermion construction, again identifying
zero-mode poles in the $t$-channel with $\gamma$ and $Z$-exchange.  As
before, we require that the Chan-Paton factors are constrained to
cancel the exotic zero modes in the other channels.  To introduce the
SM factors, we adopt the device that fermion labels in the Chan-Paton
factors are the guide to constructing the low energy limit.  This is
because the $\lambda$'s of the external legs depend on the
$SU(2)\otimes U(1)$ embedding in a larger (unifying) group, and the
Chan-Paton traces over $\lambda$'s are linked to the quantum numbers
of the $s,t$ and $u$-channels.  The connection between the string
amplitude zero mode poles and the SM poles, in keeping with this
philosophy, is described next.

We consider separately the low energy matching for
$2 \rightarrow 2$ amplitudes for (1) all left-handed ($L$)
or all right-handed ($R$); and (2) $LR \to LR$ and  $RL \to RL$.

\vskip 2mm
\underline{(1). $\ell_{\alpha}q_{\alpha} \to \ell_{\alpha}q_{\alpha};\  \alpha=L,R$ }

\vskip 2mm
The string and SM electroweak tree amplitudes for the like-helicity combinations are 
\begin{eqnarray} 
A_{string}^{NC}(\alpha\alpha) & = &ig^2
\left( S(s,t)\frac{s}{t} T_{1234} + S(t,u)\frac{s^2}{ut} T_{1324} +
S(u,s)\frac{s}{u} T_{1243} \right)\\
A_{EW}^{NC}(\alpha\alpha)  & = &2ie^2 \frac{s}{t} \left(Q_{q}Q_{\ell}+\frac{t}{t -M^2_{Z}}\ 
\frac{g^\ell_{\alpha}g^q_{\alpha}}{s^2_{W}c^2_{W}}\right)
\end{eqnarray}
where $Q_{q,\ell}$ are the electric charge of quark and lepton; 
$s_{W}=\sin\theta_{W}, c_{W}=\cos\theta_{W}.$
Matching with $e=g \sin\theta_{W}$ gives
\begin{eqnarray}
T_{1243}  & = & T_{1324} \equiv T\\
T_{1234}  & = &
T + 2 s^2_{W} \left( Q_{q}Q_{\ell} 
+\frac{t}{t-M^2_{Z}} \ \frac{g^\ell_{\alpha}g^q_{\alpha}}{s^2_{W} c^2_{W}}\right),
\end{eqnarray}
which guarantees that there is no zero-mode exotic $u$-channel pole
and that the SM tree amplitude is recovered in the limit $S(s,t) =
S(t,u) = S(s,u) \to 1$ where $s \ll M_{S}^2$.  Because the string
models have nothing to say about the electroweak symmetry breaking, we
put the effect in by hand in our treatment.  We choose to introduce
it through the condition on the Chan-Paton factors, which leads to the 
$t$-dependence in Eq.~(13). In the absence of breaking, all of the gauge
bosons would be massless $M_Z, M_W = 0$, and Eq.~(13) would relate 
the Chan-Paton factors only through the gauge charges.

 Our modified string amplitude now reads
\begin{eqnarray}
A_{string}^{NC}(\alpha\alpha) =ig^2T\frac{s}{ut}f(s,t,u)+ 2 ig^2s^2_{W} S(s,t)\frac{s}{t} \left(Q_{q}Q_{\ell}+
\frac{t}{t-M^2_{Z}}\ \frac{g^\ell_{\alpha}g^q_{\alpha}}{s^2_{W} c^2_{W}} \right),
\label{nue1}
\end{eqnarray}
where $f(s,t,u)$ was defined in Eq.~(\ref{eq:6}). 
Our convention for the SM neutral-current couplings is
\begin{eqnarray}
g^f_{L}  = T_{3f}-Q_{f}\sin^2\theta_{W}, \quad  g^f_{R}  = -Q_{f}\sin^2\theta_{W}.
\end{eqnarray}

We have adopted the shorthand that all parameters proportional to
Chan-Paton factors are designated by the single symbol $T$.  In fact,
in our study of the low energy constraints on the models in the
following section, we will make the simplifying assumption that the
factors are all equal.

\vskip 2mm
\underline{(2). $\ell_{\alpha}q_{\beta} \to \ell_{\alpha}q_{\beta};\ \alpha,\beta=L,R;\ \alpha\neq\beta$ }

\vskip 2mm
The string and SM electroweak tree amplitudes are 
\begin{eqnarray} 
A_{string}^{NC}(\alpha\beta) & = &ig^2
\left( S(s,t)\frac{u^2}{st} T_{1234} +
S(t,u)\frac{u}{t} T_{1324} +
S(u,s)\frac{u}{s} T_{1243} \right)\\
A_{EW}^{NC}(\alpha\beta) & = &2ie^2 \frac{u}{t} \left(Q_{q}Q_{\ell}+\frac{t}{t-M^2_{Z}}
\frac{g^\ell_{\alpha}g^q_{\beta}}{s^2_{W}c^2_{W}}\right).
\end{eqnarray}
Again, matching with $e=g\sin\theta_{W}$ gives
\begin{eqnarray}
T_{1243}  & = &T_{1234} \equiv T\\
T_{1324}  & = &
T + 2 s^2_{W} \left( Q_{q}Q_{\ell} 
+\frac{t}{t-M^2_{Z}} \ \frac{g^\ell_{\alpha}g^q_{\alpha}}{s^2_{W} c^2_{W}}\right).
\end{eqnarray}
The final string amplitude reads
\begin{eqnarray}
A_{string}^{NC}(\alpha\beta) = ig^2T\frac{u}{st}f(s,t,u)
+ 2 ig^2 s^2_{W} S(t,u)\frac{u}{t} \left(Q_{q}Q_{\ell} +\frac{t}{t-M^2_{Z}}
\frac{g^\ell_{\alpha}g^q_{\beta}}{s^2_{W} c^2_{W}} \right).
\label{ab}
\end{eqnarray}
This is $s\leftrightarrow u$ crossing from Eq.~(\ref{nue1}).

To obtain other amplitudes involving anti-fermions, it is a matter of simple crossing.
For example,  for Drell-Yan process $q\bar q\to \ell \bar\ell$, 
we simply have $s \leftrightarrow t$ crossing of the above formulas 
in Eqs.~(\ref{nue1}) and (\ref{ab}).

The amplitudes we have constructed are particularly convenient for comparing
to the contact interaction amplitudes analyzed and constrained by data
in the literature \cite{kingman, PDG}.  We turn next to this comparison,
deriving constraints on $M_{S}$ in the process.

\section{Linking String amplitudes to Contact Interactions}
\label{contact}

In this section, we convert constraints on contact interactions to
constraints on the string scale $M_{S}$ for given $T$ values.  In
order to compare to data at low energies, we express string deviation
from SM electroweak amplitude by $\triangle_{\alpha\beta}\ (\alpha, \beta=L, R)$, namely
\begin{equation}
A_{string}({\alpha\beta})=A_{EW}({\alpha\beta}) +\triangle_{\alpha\beta}.
\end{equation}
Using Eq.~(\ref{eq:6}), we find for like-helicity fermion scattering $(\alpha\alpha)=LL$ and $ RR$
\begin{eqnarray}
\label{delta1}
\triangle_{\alpha\alpha} & \simeq &-\frac{\pi^2}{6}\ 
\frac{st}{M^4_{S}}\ A_{EW}(\alpha\alpha) - iT g^2\frac{\pi^2}{2}\frac{s^2}{M^4_{S}},
\end{eqnarray}
where $T$ is the generic parametrized Chan-Paton factor corresponding to the
particular process.
For unlike-helicity combinations in the neutral current case, 
 $\triangle_{\alpha\beta} =  \triangle_{\alpha\alpha}(s\leftrightarrow u)$, 

The reduced amplitudes for contact interactions from physics beyond the SM
are conventionally parameterized as \cite{bchz,kingman,PDG,peskin}
\begin{equation}
\triangle M_{\alpha\beta}^{\ell q}  =  \eta_{\alpha\beta}^{\ell q} =
\epsilon \frac{4\pi}{\Lambda_{\ell q}^2}.
\label{cont}
\end{equation}
The cutoff  $\Lambda_{\ell q}$ is the mass scale at which new physics
sets in. It presumably corresponds to the mass of the heavy strongly interacting 
particles that mediate the new interaction and it is referred as the ``compositeness
scale''.  The sign factor $\epsilon = \pm1$ allows for constructive or
destructive interference between the contact interaction and the SM amplitudes.  
Typically, in the fit to a given class of interactions,
it is designated $\Lambda_{\pm}$ to distinguish between fit values
obtained with $\epsilon = \pm 1$.

The relations between the string contribution and the reduced amplitude parameterization
can be found to be, for like-helicity fermion scattering,
\begin{eqnarray}
\triangle M_{\alpha\alpha} 
& = & \frac{\triangle_{\alpha\alpha} }{i2s} \simeq
-\frac{\pi^2}{12}g^2 \frac{s}{M^4_{S}} (F+3T).
\label{eq:29}
\end{eqnarray}
For unlike-helicity fermion scattering, 
$\triangle M_{\alpha\beta} =\triangle M_{\alpha\alpha}(s\leftrightarrow  u).$
For a Drell-Yan process, which invloves with anti-fermions, we have $s \leftrightarrow t$
from Eq.~(\ref{eq:29}). The factor $F$ includes the information for chiral couplings
and it is
\begin{equation}
\nonumber
F = \left\{
\begin{array}{l}
 \frac{t}{t-M_W^2} \qquad {\rm for\ charged\ current}, \\[3mm]
\nonumber
2s^2_W \left( Q_{q}Q_{\ell} + \frac{t}{t-M^2_Z}\ \frac{g^\ell_{\alpha}g^q_{\beta}}{s^2_W c^2_W} \right)
\qquad {\rm for\  neutral\ current }\ \ell q .
\end{array}
\right.
\end{equation}

It is interesting to note that the leading stringy corrections to the SM amplitudes
as in Eq.~(\ref{delta1}) enter at dimension-8, while the standard
parameterization for four-fermion contact interactions as in Eq.~(\ref{cont}) is of
dimension-6. Due to this additional energy-dependent suppression factor $s/\mssq$,
the constraints obtained from low energy data on $\ms$ will thus
be weaker than that on $\Lambda_{\ell q}$.

In certain more complicated brane-world models, for example intersecting 
D-branes \cite{abl}, there are corrections at dimension-6
from Kaluza-Klein excitations, winding modes as well as string oscillators. 
They lead to stronger limit on the lower bound of the string scale, 
about $2-3$ TeV \cite{abl}.

\subsection{Validity of the Approximate Amplitudes}

With the above set up, we are in position to extract bounds on the
string scale from the values of parameters of contact interactions.
A global fit of contact interactions to all of 
the data discussed above plus the low energy data from neutral current
and charged current process, including atomic parity violation, is
also reported in \cite{kingman}. The low energy data dominate
these global constraints. As noted earlier, the ${s}/{M_{S}^2}$ dependence 
of our string amplitudes severely suppresses stringy effects at very low energies
and  the low energy data are insensitive to the string scale.
We will thus mainly make use of the data at highest energies 
available like in HERA, Tevatron and LEP-II.

Our expansion of the factors $S(x,y)$, where $x,y =
s,t \mbox{ or } u$, should be valid if bounds on $\ms$ are found to
be well above the kinematical region covered by the data. How close
can the scale be to the kinematical range of the data before the
approximate expansion becomes unreliable?  We address this question by
computing the CC cross section $e^- p \rightarrow \nu + X$ with the
full amplitudes and with the approximated amplitudes.  The
differential DIS cross section, in terms of the functions $V$ in
$A_{\str}^{CC}(LL)$ in Eq.~(\ref{eqV}) and $\overline{V}$ in
$A_{\str}^{CC}(L\overline{R})$  in Eq.~(\ref{eqVb}), reads
\begin{equation}
\frac{d^2\sigma}{dxdQ^2}=\frac{d\sigma^{SM}}{dQ^2} [(u(x,Q^2)+
c(x,Q^2))V^2+(1-y)^2(\overline{d}(x,Q^2)+\overline{s}(x,Q^2))\overline{V}^2],
\end{equation}
where ${d\sigma^{SM}}/{dQ^2}$ is the SM $W$-exchange Born
term differential cross section. In the course of this study, we can
probe as well the simple constraint on the model that follows from the
measured total cross section \cite{zeus1,zeus2,zeus3}, namely
\begin{displaymath}
 \sigma(Q^2 > 200\  \gev^2) =66.7 \err{3.2}{2.9}  {\rm pb},
\end{displaymath}
at $E_{CM}= 318$ GeV, the $ep$ C.M. energy.  The ZEUS collaboration quotes the value
\begin{displaymath}
\sigma(Q^2 > 200\  \gev^2) = 69.0\err{1.6}{1.3} {\rm pb}
\end{displaymath}
as the SM expectation using its NLO QCD fit.
For example, with $T = 1$  one finds the experimental 95\% CL limit
\begin{equation}
                 M_S \geq 0.45 \  \tev,
\end{equation}
whether one uses the full or the approximate amplitude.  In general
the approximate cross-sections agree with the complete calculation to
3 figures until $M_{S} \simeq E_{CM}$, where one finds differences of
the order of a percent.  For example, with $T = 1$ and $M_{S} = 320\
\gev$, the full and approximate cross sections are 85.2 pb and 82.8
pb, while with $T = -1$ the cross sections are 62.2 pb and 62.5
pb. The approximation is evidently quite good so long as $M_{S} >
E_{CM}$, since the lowest Regge resonance slips into the physical
region when $M_{S} \leq E_{CM}$ and should, in principle, be
represented by a resonant form with finite width.  However,the
vanishing of the structure functions as $x \rightarrow 1$ minimizes
the impact of the nearby resonance on the DIS cross section as $M_{S}
\rightarrow E_{CM}$ from above.

\subsection{Evaluation of Lower Limits on $M_{S}$}

Focusing on the chiral amplitudes $A_{LL}$, which enter in both the
NC and CC processes, we combine Eqs.~(\ref{cont}) and (\ref{eq:29}) to express the
constraint on $M_{S}$ at a given $T$ value and $\Lambda$ bound value as
\begin{equation}
M_S > [-\frac{\pi^2}{12}\frac{g^2s}{\eta}(F + 3T)]^{\frac{1}{4}} \mbox{ for\ DIS\ at\ HERA}.
\label{her}
\end{equation}
For the DY process at the Tevatron and $e^+e^-$ annihilations at LEP-II, 
we have $s \leftrightarrow t$
in Eq.~(\ref{her}).

In Table I we show the lower bounds on $M_{S}$ that follow from the
corresponding best fit values of $\eta$ from the HERA NC data, the
Drell-Yan data from Tevatron and the hadronic cross section from LEP-II 
quoted in \cite{kingman}. These values follow from our NC analysis
above. 
In the table we also use the NC data with the $SU(2)$ relation
between the CC and NC amplitudes, namely

\begin{displaymath}
\triangle M_{LL}(CC) = \triangle M_{LL}^{ed} - \triangle M_{LL}^{eu},
\end{displaymath}
to give corresponding limits on the CC amplitudes.  These are not
independent constraints, of course, but simply show the impact of the
data in the CC sector.   
We also include the direct CC bound on $M_{S}$ obtained in the preceding
subsection from HERA data and the DY bound obtained by CDF at the
Tevatron on the CC $qqe\nu$ compositeness scale \cite{CDF01}, with the
corresponding $M_{S}$ bound.
\begin{table}[tb]
\medskip
\centering
\begin{tabular}{|c||cc|cc|cc|cc|} \hline
 & \multicolumn{2}{c|}{HERA NC} & \multicolumn{2}{c|}{Drell-Yan}  &
 \multicolumn{2}{c|}{LEP} \\ \hline
& \underline{$\eta$ (TeV$^{-2}$)} &
 \underline{$M_S\ ({\tev})/T$} & \underline{$\eta$} & \underline{$M_S/T$}
 & \underline{$\eta$} & \underline{$M_S/T$} \\
$\eta_{LL}^{eu}$  & $-1.18 \err{0.53}{0.56}$ & $0.34/+1$  & 
$-0.19 \err{0.24}{0.21}$ & $0.85/+1$  & $-0.22 \err{0.086}{0.084}$ & $0.32/0$ \\
 &&&&&& $0.50/-1$ \\
$\eta_{LL}^{ed}$  & $1.53 \err{1.59}{1.35}$ & $0.29/-1$  & 
$ 0.88 \err{0.58}{0.73}$ & $0.34/0$  & $ 0.26 \err{0.095}{0.098}$ & $0.29/0$ \\
 &&&& $0.57/+1$  && $0.48/+1$ \\
 &&&& $0.73/-1$  &&           \\ \hline 
$\eta_{CC}$ & $ 2.71 \err{1.67}{1.46}$ & $0.26/-1$  & $ 1.07 \err{0.62}{0.76}$ & $0.41/0$  & $ 0.48 \err{0.13}{0.13}$ & $0.33/0$ \\
 &&&& $0.58/+1$  && $0.45/+1$ \\
 &&&& $0.73/-1$  &&           \\ \cline{1-3}
 & HERA $\sigma_{CC}$: & $ 0.45/+1$& (CDF) $0.80$ & $0.53/0$ &&\\
 &&&&  $0.75/+1$  &&           \\ \hline
\end{tabular}
\caption[]{\label{table1}
\small
Lower bounds on the string scale $\ms$ from contact interaction parameters,
at a $95\%$ CL.  The Chan-Paton factor $T$ has been taken as $\pm 1$ as
indicated.}
\end{table}
When translating the existing constraints on $\eta_{\alpha\beta}$ to
$\ms$, we need to take into account the different energy-dependence as
noted earlier.  In computing the values of the bounds in Table I, we
use the rule of thumb that the average parton energy fraction is
$\langle x \rangle \simeq 1/3$, so the direct channel HERA parton CM
energy squared is $s \simeq E_{CM}^2/{3}\approx (0.18\ \tev)^2$.  At
the Tevatron, where the total CM energy was $1.8\ \tev$, our nominal
parton CM energy squared is $s\simeq E_{CM}^2/{9} = (0.6\ \tev)^2$.
For the momentum transfer squared we take $Q^2 = {s}/{2}$.

In the following subsections we explain the entries in the table. 

\subsection{HERA NC}
Limits on the deviation from SM predictions for processes at HERA lead
to corresponding bounds on string parameter.  From Table IV of
\cite{kingman}, the limits on $\triangle M_{LL}$ = $\eta_{LL}^{\ell
q}$, provided separately for $eeuu$ and $eedd$, are given.  At
$2\sigma$ level (or $95\%$ CL), we have the lower bound
$\eta_{LL}^{eu} = -2.3/\tev^2$.  We apply the weak isospin constraint
that the $eeuu$ and $eedd$ amplitudes have opposite sign, which
implies the upper bound $\eta^{ed}_{LL}=4.7/\tev^2$.  In order to
obtain a lower bound on string scale $M_S$, we need the correct sign
of $\triangle M$ from our string expression corresponding to each
limit on value of $\eta$.  Consequently, in the $eeuu$ case, the gauge
factor $(F+3T) \geq 0$ is required.  In the $eedd$ case, the
requirement is $(F+3T) \leq 0$.  With typical values ${s}= (0.18
\mbox{ TeV})^2$, $t/(t-M^2_{Z})\simeq 1/2$ and $T=+1\ (-1)$, we find the
bounds $0.34\  (0.29) \mbox{ TeV}$ as shown in the table.  We should comment
here that, the typical bounds on masses of leptoquark resonances at
HERA are in the range $0.25-0.29 \mbox{ TeV}$ \cite{zeus3}, roughly
compatible with bounds from our contact interaction analysis.
Slightly higher values of $|T|$ produce higher bounds on $M_{S}$.  For
example, with $T = -2$, the value is 0.35 TeV for $eedd$.  Clearly
larger absolute values of $T$ correspond to larger bounds on $M_{S}$,
limited only by the requirement that the effective coupling constants
remain perturbative, consistent with our string amplitude
construction.  From Eq.~(\ref{her}), we see that $M_{S}\propto
(F+3T)^{1/4}$, or roughly proportional to $T^{1/4}$.  This is also the
case for DY processes at the Tevatron and $e^+e^-$ annihilation at LEP-II.

\subsection{Drell-Yan at the Tevatron}

We follow the same pattern as described above, now using
$s\leftrightarrow t$ of Eq.~(\ref{eq:29}), for limits from DY
processes at the Tevatron. For typical values we find the strongest
bounds on string scale are $0.85, 0.73$ TeV for modest values
$T=+1,-1$ for $eeuu$ and $eedd$ respectively.  An independent search
for deviations from the SM in the DY channel $qq\nu l$ at CDF
\cite{CDF01}, cited in \cite{PDG}, yields a $95\%$ CL upper bound of
$0.8/\tev^2$ on the value of $\eta_{CC}$.  The corresponding limits on
$M_{S}$ are independent of those derived from the $eedd$ case.  Searches for 
$W^\prime$ and $Z^\prime$ resonances at the Tevatron yield bounds similar to the larger of the
bounds just quoted, namely in the range $0.75 - 0.85$ TeV
\cite{CDF01}.  As in the case of leptoquark resonance searches at
HERA, the bounds on the $W^\prime$and $Z^\prime$ masses at the
Tevatron are roughly consistent with the contact interaction bounds we
just described.  The larger DY bounds rise to $0.86$ TeV and $1.04$
TeV when the $T$ values are doubled to $\pm{2}$, indicating that
increasing the magnitude of $T$ has a marked effect on $M_{S}$. In
Fig. 1 we show the plot of the lower bound on $M_{S}$ vs. the
Chan-Paton parameter $T$ in the range $1\leq |T|\leq 4$ for the $eeuu$
and $eedd$ cases, which give representative largest lower bounds on
$M_{S}$ for a given $T$ value.  In any case, it is fair to say that
the resonant bounds and the contact interaction bounds are
complementary ways to probe for string physics at the TeV scale.
  
\begin{figure}[]
\centering
\epsfxsize=4.0in
\hspace*{0in}
\epsffile{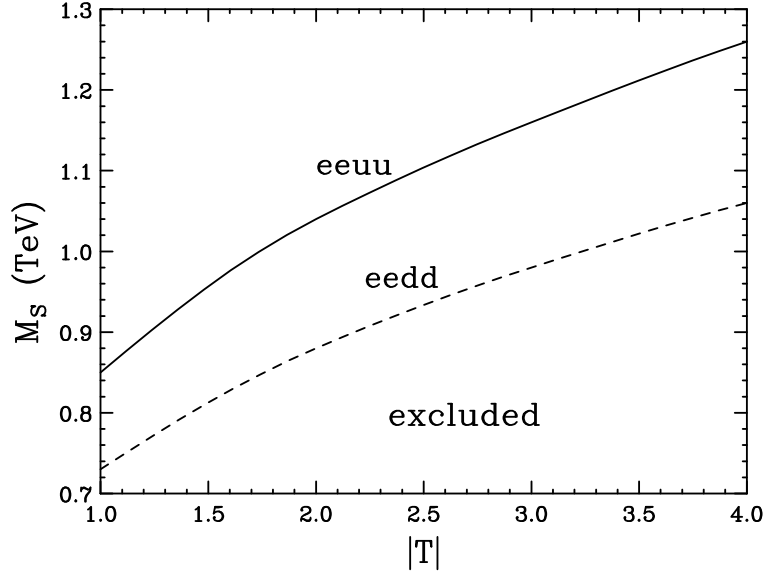}
\caption{
\label{eLeRoverM1-fig}
\small Relationship between Chan-Paton parameters and lower bounds of
the string scale $M_S$ from the DY process at the Tevatron.  $T$ is
positive and negative for $eeuu$ and $eedd$ respectively.  The region
under the curves is excluded at $95\%$ CL.  }
\end{figure}

\subsection{LEP-II}

The LEP-II results are from the lowest nominal energy, but have the
advantage that all of the CM energy can go directly into producing new
physics.  For LEP-II, we use $s = (0.2 \mbox{ TeV})^2\ \mbox{with}\
t\simeq -s/2$.  We only consider cross-section for hadron production
as stated in \cite{bchz,kingman}.  Limits are as listed in Table
I. The limits tend to be stronger than in the DIS at HERA case, since
the values of $\eta 's $ and their uncertainties are significantly
smaller and the CM energy is slightly larger than the characteristic
value used in our HERA analysis.  A consistent but somewhat weaker limit is 
given in Ref.~\cite{cpp} with $M_{S} \geq 0.41$ TeV

\section{Summary and Conclusions}
\label{conclude}
Combining the low energy limit of string amplitudes for NC and CC
processes, we find that bounds on the string scale can be obtained
that complement and extend previous analyses. In particular, we extend
previous models to cover all neutral current phenomena and, for the
first time, offer a model of charged current amplitudes in a string
resonance framework.  The essence of the approach we adopt is that of
Ref.~\cite{cim}.  The low energy limit of each string amplitude
reproduces the corresponding SM amplitude.  This leaves only a limited
number of Chan-Paton factors unspecified, and these are treated as free
parameters whose values are related by requiring consistency with the
perturbative construction of the string amplitudes.  In the absence of
new physics signals, they are constrained by the agreement between the
SM and the data for a given string scale. More generally, the
parameter space consists of the string scale $M_S$ and a limited
number of free dimensionless parameters denoted generically by $T$.
We refer to this as a ``bottom up'' approach to probing the string
aspect of braneworld.

We have focused in this paper on the match between the low energy
limit of the open-string four-fermion amplitudes at typical
kinematical region and the constraints on contact interaction
parameters determined by data from HERA, Tevatron and LEP-II.  The
bounds on the string mass scale are comparable in every case to those
found in specific models or from leptoquark and $W^\prime$ and
$Z^\prime$ searches at HERA and Tevatron.  This is no surprise, since
the accelerator energy and the precision of the measurements dictate
the accessible scale in searches for new physics.  It is also no
surprise that the highest energy data provide the highest values of
the lower bound on new physics.  The Drell-Yan processes at the
Tevatron lead to our strongest constraints, namely
\begin{equation}
\nonumber
M_S \geq \left\{
\begin{array}{l}
 0.9\ \tev\ \ \mbox{ for\ $|T|=1$}, \\[3mm]
\nonumber
 1.3\ \tev\ \ \mbox{ for\ $|T|=4$},
\end{array}
\right.
\label{stm}
\end{equation}
as shown in Fig.~1 for the $eeuu$ case.

The relationship between string scale $M_S$ and Quantum Gravity scale
$M$ is model-dependent \cite{stringy,cpp}. However, $M_{S} < M$
quite generally, so the bound on $M_{S}$ applies to $M$ as well.  
In one
simple case of D-brane scenario, the string scale and the quantum
gravity scale in the weakly coupled string sector are related by
\cite{cpp}
\begin{equation}
\frac{M}{M_{S}} = \frac{k}{g^{1/2}}
\label{qug}
\end{equation}
where the model-dependent factor $k$ is of order 1.  Taking the value
$k= 1$ and the $SU(2)$ gauge coupling at the weak scale for
illustration, we obtain from Eqs.~(\ref{stm}) and~(\ref{qug}) a
conservative bound on the gravity scale
\begin{equation}
\nonumber
M \geq \left\{
\begin{array}{l}
 1.1\ \tev\ \ \mbox{ for\ $|T|=1$}, \\[3mm]
\nonumber
 1.6\ \tev\ \ \mbox{ for\ $|T|=4$}
\end{array}
\right.
\end{equation}
from the Drell-Yan analysis of the Tevatron data. This estimate of
the range of values of the scale of gravity in large extra dimensions
is competitive with the current accelerator search values and the
value from the specific model of Ref. \cite{cpp}. But again we advise 
caution because of the model dependence of our estimate.

We conclude that a TeV string scale can measurably modify weak current
amplitudes even well below the string scale.  The corresponding limits
on this scale and the scale of gravity are quite interesting and worth
further exploration.  Including these considerations in the
interpretation of future data will add an extra dimension, or more, to
the search for new physics at the TeV scale.

\vskip 0.2cm
\noindent
{\bf Acknowledgements} 

We wish to thank K. S. Narain, G. Shiu and G. Thompson
for helpful discussions. This work was supported in part by the U.S.
Department of Energy under grant numbers 
DE-FG02-95ER40896 and DE-FG03-98ER41079, 
and in part by the Wisconsin Alumni Research Foundation.

\end{document}